\documentstyle[preprint,aps]{revtex}
\begin{document}
\title{{\bf Mass Spectra of Heavy Quarkonia and $B_{c}$ Decay Constant for Static
Scalar-Vector Interactions with Relativistic Kinematics}}
\author{Sameer M. Ikhdair\thanks{%
sameer@neu.edu.tr} and \ Ramazan Sever\thanks{%
sever@metu.edu.tr}}
\address{$^{\ast }$Department of Electrical Engineering, Near East\\
University, Nicosia, North Cyprus, Mersin 10, Turkey.\\
$^{\dagger }$Department of Physics, Middle East Technical University,\\
Ankara, Turkey.\\
PACS NUMBER(S): 11.10.St, 12.39.Pn, 14.40.-n}
\date{\today
}
\maketitle

\begin{abstract}
We reproduce masses of the self-conjugate and non-self-conjugate mesons in
the context of the spinless Salpeter equation taking into account the
relativistic kinematics and the quark spins. The hyperfine splittings for
the ${\rm 2S}$ charmonium and ${\rm 1S}$ bottomonium are also calculated.
Further, the pseudoscalar and vector decay constants of the $B_{c}$ meson
and the unperturbed radial wave function at the origin are also calculated.
We have obtained a local equation with a complete relativistic corrections
to a class of three attractive static interaction potentials of the general
form $V(r)=-Ar^{-\beta }+\kappa r^{\beta }+V_{0},$ with $\beta =1,1/2,~3/4$
decomposed into scalar and vector parts in the form $V_{V}(r)=-Ar^{-\beta
}+(1-\epsilon )\kappa r^{\beta }$ and $V_{S}(r)=\epsilon \kappa r^{\beta
}+V_{0};$ where $0\leq \epsilon \leq 1.$ We have used the shifted
large-N-expansion technique (SLNET) to solve the reduced equation for the
scalar $(\epsilon =1),$ equal mixture of scalar-vector $(\epsilon =1/2),$
and vector $(\epsilon =0)$ confinement interaction kernels. The energy
eigenvalues are carried out up to the third order approximation.
\end{abstract}

\begin{center}
$\bigskip $
\end{center}

\begin{verbatim}
 
\end{verbatim}

\section{INTRODUCTION\noindent}

The spinless Salpeter (SS) equation represents a standard approximation to
the Bethe-Salpeter equation [1]. Upon elimination of any dependence on
time-like variables in a suitable manner, the Bethe-Salpeter (BS) equation
reduces to the Salpeter equation [2]. Neglecting, furthermore, any reference
to the spin degrees of freedom and restricting to positive energy solutions,
one arrives at the SS equation, which is the correct tool to deal with the
bound-state spectrum of \ $q\overline{q},$ $Q\overline{Q}$ and $Q\overline{q}
$ or $q\overline{Q}$ (where $q=u,~d,~s$ and $Q=c,~b,~t$) interacting via
some effective potential with no spin dependence, i.e., spin-averaged data
(SAD). This equation was solved analytically and then numerically for its
bound-state energies using different techniques by many authors [3-10].

We solved the SS equation [8-10] analytically and then numerically, for the
first time, by using the shifted large-N-expansion technique (SLNET) for a
wide class of static phenomenological and QCD-motivated potentials
previously proposed for quarkonium [11-13]. The non-local SS equation was
expanded formally to order of $v^{2}/c^{2}$ up to the first relativistic
correction term. Thus, the resulting equation is a local
Schr\"{o}dinger-type equation for two constituent interacting particles of
masses $m_{q}$ and $m_{Q}.$ On the other hand, it consists of heavy quarks
and it can be reliably described by the use of the methods developed for the 
$c\overline{c}$ and the $b\overline{b}$ spectra.

The relativistic effects are not negligible in quarkonia, and we need to
investigate them. The simplest relativistic effect is just kinematics. They
are considered in the potential as nonstatic terms, including both
spin-dependent and spin-independent parts. Therefore, the nonstatic terms in
the potential, as in the Fermi-Breit approximation include a spin-spin
hyperfine interaction, a tensor interaction, a spin-orbit interaction, and a
spin-independent interaction [14-20]. There are also nonstatic terms in the
potential which are independent of spin. It is difficult to isolate these
terms from data, and furthermore, there is some controversy concerning their
functional form [21]. In this situation, it is better to omit them
altogether.

The discovery of the $B_{c}$ \ (the lowest pseudoscalar $^{1}S_{_{0}}$
state) was reported in 1998 by the collider Detector at Fermilab (CDF)\
collaboration in $1.8~TeV\ p$-$\overline{p}$\ collisions at the Fermilab $%
\left[ 22\right] $. The observed mass $M_{B_{c}}=6.40\pm 0.39\pm 0.13$ $GeV~$%
has inspired new theoretical interest in the problem [23-31]$.$ Some
preliminary estimates of the $c\overline{b}$ bound state masses have been
made devoted to the description of the charmonium and bottomonium
properties. More experimental data are expected to come in near future from
new hadron colliders.

The spectroscopy of the $c\overline{b}$ system have already been widely
studied various times over the past years in the framework of the heavy
quarkonium theory $\left[ 23\right] $. The revised analysis of the $B_{c}$
spectroscopy has been performed in the framework of the potential approach
and QCD sum rule $\left[ 24,27\right] .$ Kwong and Rosner $\left[ 28\right] $
predicted the masses of the lowest $c\overline{b}$ vector (triplet) and
pseudoscalar (singlet) states using an empirical mass formula and a
logarithmic potential. Eichten and Quigg $\left[ 22\right] $ gave a more
comprehensive account of the energies and properties of the $c\overline{b}$
system that was based on the QCD-motivated potential of Buchm\"{u}ller and
Tye [32]. Gershtein {\it et al.} [29] also published a detailed account of
the energies and decays of the $c\overline{b}$ system using a QCD sum-rule
calculation. Baldicchi and Prosperi [27] have computed the $c\overline{b}$
spectrum based on an effective mass operator with full relativistic
kinematics. They have also fitted the entire quarkonium spectrum. Fulcher 
{\it et al. }[25] extended the treatment of the spin-dependent potentials to
the full radiative one-loop level and thus included effects of the running
coupling constant in these potentials. They have also used the
renormalization scheme developed by Gupta and Radford [32]$.$ We applied the
nonrelativistic form of the statistical model [30] and shifted large-N
expansion technique (SLNET) [31] to calculate the spectroscopy, decay
constant and other properties of the heavy mesons, including the $c\overline{%
b}$ system.

Recently, in 2002, the ALEPH collaboration has searched for the pseudoscalar
bottomonium meson, the $\eta _{b}$ in two-photon interactions at LEP2 with
an integrated luminosity of $699$ pb$^{-1}$ collected at $e^{+}e^{-}$
centre-of mass energies from $181$ ${\rm GeV}$ to 209 ${\rm GeV.}$ One
candidate event is found in the six-charged-particle final state and none in
the four-charged-particle final state. The candidate $\eta _{b}$ ($\eta
_{b}\rightarrow K_{S}K^{-}\pi ^{+}\pi ^{-}\pi ^{+}$) has reconstructed
invariant mass of $9.30\pm 0.02\pm 0.02$ ${\rm GeV}$ [33]${\rm .}$
Theoretical estimates (from perturbative QCD and lattice nonrelativistic QCD
of the mass splitting between $\eta _{b}({\rm 1S)}$ and $\Upsilon ({\rm 1S}%
), $ $M$($\Upsilon ({\rm 1}^{3}{\rm S}_{1}))=9.460$ ${\rm GeV,}$ are
reported (cf. [33] and references therein).

Further, in 2002, the Belle Collaboration [34] has observed a new
pseudoscalar charmonium state, the $\eta _{c}({\rm 2S}),$ in exclusive $%
B\longrightarrow KK_{S}K^{-}\pi ^{+}$ decays. The measured mass of the $\eta
_{c}({\rm 2S}),$ $M$($\eta _{c}({\rm 2S}))=3654\pm 14$ ${\rm MeV.}$ It is
close to the $\eta _{c}({\rm 2S})$ mass observed by the same group in the
experiment $e^{+}e^{-}\longrightarrow J/\psi \eta _{c}$ where $M$($\eta _{c}(%
{\rm 2S}))=3622\pm 12$ ${\rm MeV}$ was found [35]. It is giving rise to a
small hyperfine splitting for the ${\rm 2S}$ state, $\Delta _{{\rm hfs}}(%
{\rm 2S,}$exp)$=M({\rm 2}^{3}{\rm S}_{1})-M({\rm 2}^{1}{\rm S}_{0})=32\pm 14$
${\rm MeV}$ [36]${\rm .}$ Badalian and Bakker [37] calculated the hyperfine
splitting for the ${\rm 2S}$ charmonium state in a recent work. Recksiegel
and Sumino developed a new formalism [38] based on perturbative QCD to
compute the hyperfine splittings of the bottomonium spectrum as well as the
fine and hyperfine splittings of the charmonium spectrum [39].

The motivation of the present calculation is to extend the SLNET
[8-10,31,40-42] to the treatment of the SS equation [10] to calculate the
hyperfine splittings for the ${\rm 2S}$ charmonium and ${\rm 1S}$
bottomonium and also to reproduce the $c\overline{b}$ Salpeter binding
masses below the continuum threshold $M(4S).$ We shall use a potential model
that includes a running coupling constant effects in the central potential
to give a simultaneous account of the properties of the $c\overline{b}$
system. We choose a class of three static central potentials [10-13$,$%
30,31,41,42] each with a strong coupling constant $\alpha _{s}$ to fit the
spectroscopy of the existing and non-existing quarkonium systems. Since one
would expect the average values of the momentum transfer in the various
quark-antiquark states to be different, some variation in the values of the
strong coupling constant and the normalization scale in the spin-dependent
should be expected. We extend our previous work [31] for the solution of the
SS equation to determine the binding masses of the $c\overline{c},~b%
\overline{b},~$and $c\overline{b}$ mesons by taking into account the
spin-spin, spin-orbit and tensor interactions [14-21]. The spin effects are
treated, in the framework of perturbation theory, as perturbation to the
static potential and the treatments are based on the reduced Salpeter
equation.

The outline of this paper is as following: In Section II, we first review
briefly the analytic solution of the SS equation, we consider here the
unequal mass case studied in our earlier works for a class of static
potentials using SLNET. Section III is devoted for the class of three static
potentials, which are decomposed into scalar and vector parts and also for
their spin corrections. Pure vector and scalar potentials as well as their
mixture are considered. Section IV is devoted for the pseudoscalar and
vector leptonic constant of the $B_{c}$ meson. We finally present an
approximation to $c\overline{c},~b\overline{b},~$and $c\overline{b}$ mass
spectra and decay constant of the $B_{c}$ meson. We also calculate the the
observed new charmonium $\eta _{c}{\rm (2S)}$ and searched bottomonium $\eta
_{c}{\rm (1S)}$ mesons and the hyperfine splittings of their states.
Finally, our conclusions are given in Section V. Appendix A contains some
definitions as well as the formulas necessary to carry out the above
mentioned computations.\ \ \ \ \ \ \ \ \ \ \ \ \ \ \ \ \ \ \ \ \ \ \ \ \ \ \
\ \ \ \ \ \ \ \ \ \ \ \ \ \ 

\section{WAVE EQUATION}

The relativistic wave Salpeter equation [8-10] is constructed by considering
the kinetic energies of the constituents and the interaction potential. We
assume that the full Hamiltonian $H$ governing the dynamics of the
quantum-mechanical system under consideration can be split up into a free
Hamiltonian $H_{0}$ and an interaction potential $V(r).$ For the case of two
particles with unequal masses $m_{q}$ and $m_{Q},$ interacting via a
spherically symmetric potential $V(r),$ the spatial coordinate
representation of the SS equation in the center-of-momentum system for the
two-body system reads off

\begin{equation}
\left[ 
%TCIMACRO{\underset{i=q,Q}{\sum }}%
%BeginExpansion
\mathrel{\mathop{\sum }\limits_{i=q,Q}}%
%EndExpansion
\sqrt{-\Delta _{N}+m_{i}^{2}}+V(r)-M_{n,l}(q\overline{Q})\right] \Psi \left( 
{\bf r}\right) =0,  \label{1}
\end{equation}
with the potential $V(r)$ contains, in addition to the static interaction $%
V_{static}(r)$ the total spin-dependent potential $V_{SD}(r)~$[15-18,20,21].
The kinetic terms involving the operation $\sqrt{-\Delta _{N}+m_{i}^{2}}$
are nonlocal operators and can be solved through a direct expansion of the
square root-operator [5,10,17]. For heavy quarks, the kinetic energy
operators in Eq. (1) can be approximated by the standard expansion in the
inverse powers of quark masses to obtain the usual nonrelativistic
Schr\"{o}dinger Hamiltonian [5,10]

\begin{equation}
%TCIMACRO{\underset{i=q,Q}{\sum }}%
%BeginExpansion
\mathrel{\mathop{\sum }\limits_{i=q,Q}}%
%EndExpansion
\sqrt{-\Delta _{N}+m_{i}^{2}}=m_{q}+m_{Q}-\frac{\Delta _{N}}{2\mu }-\frac{%
\Delta _{N}^{2}}{8\eta ^{3}}-\cdots ,  \label{2}
\end{equation}
where $\mu =\frac{m_{q}m_{Q}}{m_{q}+m_{Q}}$ denotes the reduced mass and $%
\eta =\mu \left( \frac{m_{q}m_{Q}}{m_{q}m_{Q}-3\mu ^{2}}\right) ^{1/3}$ is a
useful mass parameter. The radial part of Eq. (1), in the ${\rm N}$%
-dimensional space, is expanded in powers of \ $v^{2}/c^{2}$ up to one
relativistic correction term as [5,8-10,16]

\begin{equation}
\left\{ -\frac{\Delta _{N}}{2\mu }-\frac{\Delta _{N}^{2}}{8\eta ^{3}}%
+V(r)\right\} R_{n,l}(r)=E_{n,l}R_{n,l}(r),  \label{3}
\end{equation}
where $E_{n,l}=M_{n,l}(q\overline{Q})-m_{q}-m_{Q}$ stands for the Salpeter
binding energy and $\Delta _{N}=\nabla _{N}^{2}.$ \footnote{%
This approximation is correct to $O(v^{2}/c^{2}).$ The $\Delta _{N}^{2}$
term in (3) should be properly treated as a perturbation by using trial
wavefunctions [43].} This SS-type equation retains its relativistic
kinematics and is suitable for describing the spin-averaged spectrum of two
bound quarks of masses $m_{q}$ and $m_{Q}$ and total binding meson mass
eigenstate $M_{n,l}(q\overline{Q}).$ Furthermore, in order to obtain a
Schr\"{o}dinger-like equation, the perturbed term in Eq. (3) is treated
using the reduced Schr\"{o}dinger equation [44] 
\begin{equation}
p^{4}=4\mu ^{2}\left[ E_{n,l}-V(r)\right] ^{2},  \label{4}
\end{equation}
where $p$ is the center of mass momentum of the $q\overline{Q}$ system.
Consequently one would reduce Eq. (3) to the Schr\"{o}dinger-type form [17]

\begin{equation}
\left\{ -\frac{\Delta _{N}}{2\mu }-\frac{\mu ^{2}}{2\eta ^{3}}\left[
E_{n,l}^{2}+V^{2}(r)-2E_{n,l}V(r)\right] +V(r)\right\} R_{n,l}({\bf r}%
)=E_{n,l}R_{n,l}({\bf r}).  \label{5}
\end{equation}
After employing the following transformation\ \ \ 
\begin{equation}
R_{n,l}(r)=\frac{u(r)}{r^{\left( N-1\right) /2}},  \label{6}
\end{equation}
we may rewrite Eq.(5) simply as 
\begin{equation}
\left[ -\frac{1}{2\mu }\frac{d^{2}}{dr^{2}}+\frac{\left[ \overline{k}-\left(
1-a\right) \right] \left[ \overline{k}-\left( 3-a\right) \right] }{8\mu r^{2}%
}+W_{n,l}(r)-\frac{W_{n,l}(r)^{2}}{2m%
%TCIMACRO{\UNICODE[m]{0xb4}}%
%BeginExpansion
{\acute{}}%
%EndExpansion
}\right] u_{n,l}(r)=0,  \label{7}
\end{equation}
with 
\begin{equation}
W_{n,l}(r)=V_{eff}(r)-E_{n,l},  \label{8}
\end{equation}
and the effective mass parameter 
\begin{equation}
m^{^{\prime }}=\eta ^{3}/\mu ^{2}=(m_{q}m_{Q}\mu )/(m_{q}m_{Q}-3\mu ^{2}).\ 
\label{9}
\end{equation}
The perturbation term, $W_{n,l}(r)^{2},$ in (7) is significant only where it
is small (i.e., $W_{n,l}(r)/m^{^{\prime }}\ll 1)$. This condition is
verified by the confining potentials used to describe heavy$-$quark systems
except near the color$-$Coulomb singularity at the origin, and for $%
r\rightarrow \infty $. However, it is always satisfied on the average as
stated by Durand et al. [17]. We now proceed to solve Eq. (7) with Eq. (8).
In the SLNET [8-10,30,40-42], it is convenient to shift the origin of
coordinates to $r=r_{0}$ $($or $y=0)$ by defining

\begin{equation}
y=\bar{k}^{1/2}(r-r_{0})/r_{0}.  \label{10}
\end{equation}
Expansions about this point in powers of $y$ and $\overline{k}$ yield 
\begin{equation}
\frac{1}{r^{2}}=\frac{1}{r_{0}^{2}}\stackrel{\infty }{%
%TCIMACRO{\underset{j=0}{\sum }}%
%BeginExpansion
\mathrel{\mathop{\sum }\limits_{j=0}}%
%EndExpansion
}(-y)^{j}\frac{(1+j)}{\overline{k}^{-j/2}},  \label{11}
\end{equation}
\begin{equation}
V(y(r_{0}))\;=\frac{1}{Q}\stackrel{\infty }{%
%TCIMACRO{\underset{j=0}{\sum }}%
%BeginExpansion
\mathrel{\mathop{\sum }\limits_{j=0}}%
%EndExpansion
}\left( \frac{d^{j}V(r_{0})}{dr_{0}^{j}}\right) \frac{\left( r_{0}y\right)
^{j}}{j!}\overline{k}^{(4-j)/2},  \label{12}
\end{equation}
and also 
\begin{equation}
E_{n,l}\;=\frac{1}{Q}\stackrel{\infty }{%
%TCIMACRO{\underset{j=0}{\sum }}%
%BeginExpansion
\mathrel{\mathop{\sum }\limits_{j=0}}%
%EndExpansion
}\overline{k}^{(2-j)}E_{j}.  \label{13}
\end{equation}
Further, by substituting Eqs. (8) through (13) into Eq. (7), one gets

\[
\left\{ -\frac{1}{4\mu }\frac{d^{2}}{dy^{2}}+\left[ \frac{\overline{k}}{%
16\mu }-\frac{(2-a)}{8\mu }+\frac{(1-a)(3-a)}{16\mu \overline{k}}\right]
\times \stackrel{\infty }{%
%TCIMACRO{\underset{j=0}{\sum }}%
%BeginExpansion
\mathrel{\mathop{\sum }\limits_{j=0}}%
%EndExpansion
}(-y)^{j}\frac{\left( 1+j\right) }{\overline{k}^{j/2}}\right. 
\]

\[
+\frac{r_{0}^{2}\;}{Q}\stackrel{\infty }{%
%TCIMACRO{\underset{j=0}{\sum }}%
%BeginExpansion
\mathrel{\mathop{\sum }\limits_{j=0}}%
%EndExpansion
}\left( \frac{d^{j}V(r_{0})}{dr_{0}^{j}}\right) \frac{\left( r_{0}y\right)
^{j}}{j!}\overline{k}^{(2-j)/2}-\frac{r_{0}^{2}\;\overline{k}}{m^{\prime }Q}%
\left[ \stackrel{\infty }{%
%TCIMACRO{\underset{j=0}{\sum }}%
%BeginExpansion
\mathrel{\mathop{\sum }\limits_{j=0}}%
%EndExpansion
}\left( \frac{d^{j}V(r_{0})}{dr_{0}^{j}}\right) \frac{\left( r_{0}y\right)
^{j}}{j!}\overline{k}^{-j/2}\right] ^{2} 
\]
\begin{equation}
+\left. \frac{2r_{0}^{2}\;}{m^{\prime }Q}\stackrel{\infty }{%
%TCIMACRO{\underset{j=0}{\sum }}%
%BeginExpansion
\mathrel{\mathop{\sum }\limits_{j=0}}%
%EndExpansion
}\overline{k}^{(1-j)}E_{j}\times \stackrel{\infty }{%
%TCIMACRO{\underset{j=0}{\sum }}%
%BeginExpansion
\mathrel{\mathop{\sum }\limits_{j=0}}%
%EndExpansion
}\left( \frac{d^{j}V(r_{0})}{dr_{0}^{j}}\right) \frac{\left( r_{0}y\right)
^{j}}{j!}\overline{k}^{-j/2}\right\} \varphi _{n_{r}}(y)={\cal E}%
_{n_{r}}\varphi _{n_{r}}(y),\newline
\label{14}
\end{equation}
with the eigenvalues 
\[
{\cal E}_{n_{r}}=\frac{r_{0}^{2}}{Q}\left\{ \bar{k}\left( E_{0}+\frac{%
E_{0}^{2}}{m^{^{\prime }}}\right) +\left( E_{1}+\frac{2E_{0}E_{1}}{%
m^{^{\prime }}}\right) +\left( E_{2}+\frac{2E_{0}E_{2}}{m^{^{\prime }}}+%
\frac{E_{1}^{2}}{m^{^{\prime }}}\right) \frac{1}{\bar{k}}\right. 
\]

\begin{equation}
\;+\left. \left( E_{3}+\frac{2E_{0}E_{3}}{m^{^{\prime }}}+\frac{2E_{1}E_{2}}{%
m^{^{\prime }}}\right) \frac{1}{\bar{k}^{2}}+\cdots \right\} .~~~~~~~~~~~~~~
\label{15}
\end{equation}
The parameter $Q$ is an arbitrary scale, but is to be set equal $\overline{k}%
^{2}$ at the end of our calculations. Thus comparing Eqs. (14) and (15) with
its counterpart Schr{\"{o}}dinger$-$like equation for the one$-$dimensional
anharmonic oscillator problem [40], we calculate all the relevant quantities 
$\delta $'s and $\varepsilon $'s [31]. The final analytic expression in the $%
1/\bar{k}$ expansion of the energy eigenvalues appropriate to the SS
particle is

\[
{\cal E}_{n_{r}}=\bar{k}\left[ \frac{1}{16\mu }+\frac{r_{0}^{2}V(r_{0})}{Q}-%
\frac{r_{0}^{2}V(r_{0})^{2}}{m^{^{\prime }}Q}+\frac{2r_{0}^{2}E_{0}V(r_{0})}{%
m^{^{\prime }}Q}\right] 
\]

\[
+\left[ (1+2n_{r})\frac{\omega }{2}-\frac{(2-a)}{8\mu }\right] +\frac{1}{%
\bar{k}}\left[ \frac{2r_{0}^{2}E_{2}V(r_{0})}{m^{^{\prime }}Q}+\beta ^{(1)}%
\right] 
\]

\begin{equation}
+\frac{1}{\bar{k}^{2}}\left[ \frac{2r_{0}^{2}E_{3}V(r_{0})}{m^{^{\prime }}Q}%
+\beta ^{(2)}\right] +O\left[ \frac{1}{\bar{k}^{3}}\right] ,  \label{16}
\end{equation}
with $n_{r}$ is to be set equal to $0,1,2,\cdots .$ The quantities $\beta
^{(1)}$ and $\beta ^{(2)}$ appearing in the correction to the leading order
of the energy expression are displayed in [31].

Comparing the terms of Eq. (15) with their counterparts in Eq. (16) and
equating terms of same order in $\overline{k},$ one gets the leading
contribution to the binding energy

\begin{equation}
E_{0}=V(r_{0})+\frac{m^{^{\prime }}}{2}\left[ \sqrt{1+\frac{Q}{%
4r_{0}^{2}m_{\mu }m^{^{\prime }2}}}-1\right] ,  \label{17}
\end{equation}
where m$_{\mu }$=$\mu /m^{^{\prime }}.$ Here $r_{0}$ is chosen to minimize $%
E_{0},$ that is,

\begin{equation}
\frac{dE_{0}}{dr_{0}}=0{~;~~}\frac{d^{2}E_{0}}{dr_{0}^{2}}>0,  \label{18}
\end{equation}
and it satisfies 
\begin{equation}
r_{0}^{3}V^{\prime }(r_{0}){\left( \frac{m^{^{\prime }2}}{4}+\frac{Q}{%
16r_{0}^{2}m_{\mu }}\right) }^{1/2}=\frac{Q}{16m_{\mu }}.  \label{19}
\end{equation}
To solve for the shifting parameter $a$, the next contribution to the energy
eigenvalue, in Eq. (13), is chosen to vanish (i.e., $E_{1}=0)$ which gives

\begin{equation}
a=2-4\mu (1+2n_{r})\omega ,  \label{20}
\end{equation}
with $\omega $ defined as

\begin{equation}
\omega =\frac{1}{4\mu }{\left[ 3+r_{0}V^{\prime \prime }(r_{0})/V^{\prime
}(r_{0})-16r_{0}^{4}m_{\mu }V^{\prime }(r_{0})^{2}/Q\right] }^{1/2}.
\label{21}
\end{equation}
The scaling parameter $Q,$ in Eq. (19), is simply written as 
\begin{equation}
Q=8m_{\mu }{\left[ \;r_{0}^{2}V^{\prime }(r_{0})\right] }^{2}(1+\xi ),
\label{22}
\end{equation}
with

\begin{equation}
\xi =\sqrt{1+\left( \frac{{m^{^{\prime }}}}{{r_{0}V^{\prime }(r_{0})}}%
\right) ^{2}\;}.  \label{23}
\end{equation}
Therefore, with the help of relations (20) through (23) together with $Q=%
\overline{k}^{2}$, we obtain the following formula

\begin{equation}
1+2l+4\mu (2n_{r}+1)\omega =2r_{0}^{2}V^{\prime }(r_{0})\left( 2m_{\mu
}+2m_{\mu }\xi \right) ^{1/2},  \label{24}
\end{equation}
which is an explicit equation in $r_{0}$. Once $r_{0}$ is determined via
(24), it becomes easy and straightforward to obtain $E_{0}$ via Eq. (17), $%
E_{2}$ and $E_{3\text{ }}$ via \ solving Eqs. (15) -(16). Thus, the general
expression for the binding energy eigenvalue takes

\begin{equation}
E_{n,l}=E_{0}+\frac{1}{r_{0}^{2}\left( 1-\frac{2W_{n,l}(r_{0})}{m^{^{\prime
}}}\right) }\left[ \beta ^{(1)}+\frac{\beta ^{(2)}}{\bar{k}}+O\left( \frac{1%
}{\bar{k}^{2}}\right) \right] ,  \label{25}
\end{equation}
which works well and is convergent as the value of $l~$increases. Finally,
the binding meson mass eigenstate for the quarkonium families is 
\begin{equation}
M_{n,l}(q\overline{Q})=2E_{n,l}+m_{q}+m_{Q}.  \label{26}
\end{equation}
where $m_{1}$ and $m_{2}$ are constituent quark masses.

\section{HEAVY QUARKONIUM AND $B_{c}$ MESON MASS SPECTRA\noindent}

To describe the spin-dependent relativistic corrections to the potential, we
have used a similar approach derived by Olsson {\it et al. }[14] for
positron. We choose the potential in Eq. (1) as [18,20,44] 
\begin{equation}
V(r)=V_{static}(r)+V_{SD}(r)+V_{SI}(r),  \label{27}
\end{equation}
with spin-dependent and spin-independent perturbation terms are given in
Refs. [20,21,43,44]. Further, the static potential [7-13,30,31,41,42] takes
the general form

\begin{equation}
V_{static}(r)=-\frac{A}{r^{\beta }}+\kappa r^{\beta }+V_{0};~\beta
=1,1/2,~3/4,  \label{28}
\end{equation}
where $A>0,~\kappa >0$ and $V_{0}$ may be of either sign. The first
potential we consider here is the Cornell [13] potential which is one of the
earliest QCD-motivated potentials having the form

\begin{equation}
V_{C}(r)=-\frac{A}{r}+\kappa r+V_{0},  \label{29}
\end{equation}
where $\ A=4\alpha _{s}/3,$ the Coulomb-type piece characterizes the
short-range gluon exchange, $\kappa ~$is a confinement constant, and the
constant $V_{0}$ is for spin-independent interactions not included
explicitly in the $r-$ dependent part of the potential. It is related to the
slope $\kappa $ of the linear potential by [15] 
\begin{equation}
V_{0}=-2\sqrt{\kappa }\exp (-\gamma _{E}+\frac{1}{2}),  \label{30}
\end{equation}
where $\gamma _{E}=0.577215\cdots $ is the Euler-Mascheroni constant. The
second potential is that of Song and Lin [12] and is given by 
\begin{equation}
V_{S-L}(r)=-\frac{A}{r^{1/2}}+\kappa r^{1/2}+V_{0}.  \label{31}
\end{equation}
The third potential is an intermediate case between the last two mentioned
potentials and is called Turin potential [11]

\begin{equation}
V_{T}(r)=-\frac{A}{r^{3/4}}+\kappa r^{3/4}+V_{0}.  \label{32}
\end{equation}
The class of static potentials in Eq. (28) must satisfy the following
conditions

\begin{equation}
\frac{dV}{dr}>0,~\frac{d^{2}V}{dr^{2}}\leq 0.  \label{33}
\end{equation}
The fine and hyperfine splittings are computed by using the Breit-Fermi
interaction with the assumptionp [16,18]

\begin{equation}
V_{S}(r)=\epsilon \kappa r^{\beta }+V_{0},  \label{34}
\end{equation}

\begin{equation}
V_{V}(r)=V_{static}(r)-V_{S}(r),  \label{35}
\end{equation}
where $\epsilon $ is some mixing parameter. The vector term incorporates the
expected short-distance behaviour from single-gluon exchange. We have also
included a multiple of the long-range interaction in $V_{V}(r)$ to see if we
can determine the vector-scalar nature of the confining interaction. Here we
investigate the cases of pure scalar confinement $(\epsilon =1),$ equal
mixture of scalar-vector couplings $(\epsilon =1/2)$ and a pure vector case $%
(\epsilon =0).$ The spin dependent correction to the nonrelativistic
Hamiltonian, which is responsible for the fine splittings, is also modelled
on the Breit-Fermi Hamiltonian [22,29]. It can be decomposed into a part,
which is antisymmetric with respect to the spins of the constituents $%
V_{A}(r)$, and apart symmetric in these spins. The symmetric part can be
decomposed into a spin-orbit interaction $V_{LS}(r),$ a spin-spin
interaction $V_{SS}(r),$ and a tensor part $V_{T}(r).$ Therefore, it is
given by [15-17,20,21,44]

\[
V_{SD}(r)=V_{A}+V_{S}=\frac{1}{4}\left[ \frac{1}{m_{q}^{2}}-\frac{1}{%
m_{Q}^{2}}\right] \left[ \frac{V_{V}^{\prime }(r)-V_{S}^{\prime }(r)}{r}%
\right] {\bf L}\cdot {\bf S}_{-} 
\]

\[
+\frac{{\bf L}\cdot {\bf S}}{m_{q}m_{Q}}\frac{V_{V}^{\prime }(r)}{r}+\frac{1%
}{2}\left[ \frac{{\bf L}\cdot {\bf S}_{1}}{m_{q}^{2}}+\frac{{\bf L}\cdot 
{\bf S}_{2}}{m_{Q}^{2}}\right] \left[ \frac{V_{V}^{\prime }(r)-V_{S}^{\prime
}(r)}{r}\right] 
\]
\begin{equation}
+\frac{2}{3}\frac{{\bf S}_{1}\cdot {\bf S}_{2}}{m_{q}m_{Q}}\left[ \nabla
^{2}V_{V}(r)\right] +\frac{S_{12}}{m_{q}m_{Q}}\left[ -V_{V}^{\prime \prime
}(r)+\frac{V_{V}^{\prime }(r)}{r}\right] ,  \label{36}
\end{equation}
where ${\bf S}_{1}$ and ${\bf S}_{2}$ are the quark spins, ${\bf S}_{-}={\bf %
S}_{1}-{\bf S}_{2},$ ${\bf L=x\times p}$ is the relative orbital angular
momentum, and $S_{12}=T-\left( {\bf S}_{1}\cdot {\bf S}_{2}\right) /3$ where 
$T=({\bf S}_{1}\cdot \widehat{{\bf r}})({\bf S}_{2}\cdot \widehat{{\bf r}})$
\ is the tensor operator with the versor $\widehat{{\bf r}}={\bf r/}r$ . The
spin dependent correction (36), which is responsible for the hyperfine
splitting of the mass levels, in the short-range is generally used in the
form for $S$-wave $(L=0)$ (cf. e.g., [32,43,44]) 
\begin{equation}
V_{HFS}(r)=\frac{2}{3}({\bf S}_{1}\cdot {\bf S}_{2})\nabla ^{2}\left[ -\frac{%
4\alpha _{s}}{3r^{\beta }}\right] ,  \label{37}
\end{equation}
but the one responsible for the fine splittings is used for $P$- and $D$%
-waves $(L\neq 0)$ it is given by

\[
V_{FS}(r)=\frac{1}{m_{q}m_{Q}}\left\{ \frac{{\bf L}\cdot {\bf S}}{r}\left[
\left( 1+\frac{1}{4}\frac{m_{q}^{2}+m_{Q}^{2}}{m_{q}m_{Q}}\right)
V_{V}^{\prime }(r)-\frac{1}{4}\frac{m_{q}^{2}+m_{Q}^{2}}{m_{q}m_{Q}}%
V_{S}^{\prime }(r)\right] \right. 
\]

\begin{equation}
+\frac{2}{3}({\bf S}_{1}\cdot {\bf S}_{2})\nabla ^{2}\left[ \kappa
(1-\epsilon )r^{\beta }\right] +\left. \left[ T-\frac{1}{3}\left( {\bf S}%
_{1}\cdot {\bf S}_{2}\right) \right] \left[ V_{V}^{\prime \prime }(r)+\frac{%
V_{V}^{\prime }(r)}{r}\right] \right\} ,  \label{38}
\end{equation}
where the matrix element can be evaluated in terms of the expectation values
[22] $\left\langle {\bf L}\cdot {\bf S}_{1}\right\rangle =\left\langle {\bf L%
}\cdot {\bf S}_{2}\right\rangle =\frac{1}{2}\left\langle {\bf L}\cdot {\bf S}%
\right\rangle $. Hence, Eq. (38) is the complete spin-dependent potential in
QCD through order $m^{2}.$ For bound state constituents of spin $%
S_{1}=S_{2}=1/2,$ the scalar product of their spins ${\bf S}_{1}\cdot {\bf S}%
_{2}$ and ${\bf L}\cdot {\bf S}$ are to be found in the Appendix A. The
appearance of a Coulomb-like contribution $\sim 1/r$ in the vector part of
the potential causes some problems due to the relation $\nabla
^{2}(1/r)=-4\pi \delta ^{(3)}(x),$ in the spin-spin interaction (37)
involves a delta function of the $S$-wave $(L=0).$ Thus, for Cornell
potential, the hyperfine splitting potential (37) gives 
\begin{equation}
V_{eff}(r)=-\frac{A}{r}+\kappa r+\frac{32\pi \alpha _{s}}{9m_{q}m_{Q}}\delta
^{(3)}({\bf r})({\bf S}_{1}\cdot {\bf S}_{2})+V_{0};\text{ where }\beta =1.
\label{39}
\end{equation}
Therefore for the energy of spin-spin interaction we have approximately:

\begin{equation}
E_{ss}=\frac{1}{2M_{n,0}}\Delta M_{ss}^{2}\left\langle {\bf S}_{1}\cdot {\bf %
S}_{2}\right\rangle ,  \label{40}
\end{equation}
where $M_{n,0}$ is given in Eq. (26) and the singlet-triplet mass squared
difference [15] 
\begin{equation}
\Delta M_{ss}^{2}=M_{S=1}^{2}-M_{S=0}^{2}\simeq \frac{32}{9}\alpha
_{s}\kappa ,  \label{41}
\end{equation}
for light $q\overline{q}$ systems (in the instantaneous-limit approximation)
[15], and 
\begin{equation}
\Delta M_{ss}^{2}=M_{S=1}^{2}-M_{S=0}^{2}\simeq \frac{256}{3\pi ^{2}}\alpha
_{s}\kappa ,  \label{42}
\end{equation}
for heavy quarkonia (hydrogen-like trial functions) [15]. All these
predictions for the mass-squared difference are independent of the mass of
the particles which constitute the bound state. Further, for the Song-Lin
and Turin potentials, it also give

\begin{equation}
V_{eff}(r)=-\frac{A}{r^{\beta }}+\kappa r^{\beta }+\frac{8\beta (1-\beta
)\pi \alpha _{s}}{9m_{q}m_{Q}r^{2}}r^{-\beta }{\bf S}_{1}\cdot {\bf S}%
_{2}+V_{0};\text{ where }\beta =1/2,3/4.  \label{43}
\end{equation}
On the other hand, the Eq. (38), for $P$, $D,\cdots $ waves $(L\neq 0)$
case, gives

\begin{equation}
V_{eff}(r)=V_{static}(r)+g(r)\left[ F_{LS_{-}}\left( {\bf L}\cdot {\bf S}%
_{-}\right) +F_{LS}\left( {\bf L}\cdot {\bf S}\right) +F_{SS}\left( {\bf S}%
_{1}\cdot {\bf S}_{2}\right) +F_{T}T\right] ,  \label{44}
\end{equation}
with a given set of spin-dependent quantities 
\begin{equation}
F_{LS_{-}}=\left[ \frac{1}{4}\frac{m_{Q}^{2}-m_{q}^{2}}{m_{q}m_{Q}}\left[
Ar^{-\beta }+\left( 1-\epsilon \right) \kappa r^{\beta }\right] \right] ,
\label{45}
\end{equation}

\begin{equation}
F_{LS}=\left[ \left( 1+\frac{1}{4}\frac{m_{q}^{2}+m_{Q}^{2}}{m_{q}m_{Q}}%
\right) \left[ Ar^{-\beta }+\left( 1-2\epsilon \right) \kappa r^{\beta }%
\right] +\epsilon \kappa r^{\beta }\right] ,  \label{46}
\end{equation}

\begin{equation}
F_{SS}=\left[ -\frac{(2+\beta )}{3}Ar^{-\beta }+\beta \left( 1-\epsilon
\right) \kappa r^{\beta }\right] ,  \label{47}
\end{equation}
and

\begin{equation}
F_{T}=\left[ (2+\beta )Ar^{-\beta }+(2-\beta )\left( 1-\epsilon \right)
\kappa r^{\beta }\right] ,  \label{48}
\end{equation}
where $g(r)=\frac{\beta }{m_{q}m_{Q}r^{2}}$ is a necessary coupling function$%
.$ The spin-independent corrections in Eq. (27) are explicitly given by Ref.
[44] which are not treated in our present work.

\subsection{Singlet states}

For parastates $\left( L=J\right) $ or $(S=0)$ case, we have parity $%
P=(-1)^{J+1}$ and charge conjugation $C=(-1)^{L}.$ Thus, the potential (44)
can be rewritten as

\begin{equation}
V_{eff}(r)=V_{static}(r)-\frac{1}{4}\left( 3F_{SS}+F_{T}\right) +\sqrt{\frac{%
1}{10}(2L+3)(2L-1)}F_{LS_{-}},  \label{49}
\end{equation}
which can be substituted in Eq. (8) and also by setting $\overline{k}=N+2J-a$
therein$.$ Further, Eqs. (39) and (43) give 
\begin{equation}
V_{eff}(r)=-\frac{A}{r}+\kappa r-\frac{8\pi \alpha _{s}}{3m_{q}m_{Q}}\delta
^{(3)}({\bf r})+V_{0},  \label{50}
\end{equation}
and 
\begin{equation}
V_{eff}(r)=-\frac{A}{r^{\beta }}+\kappa r^{\beta }-\frac{2\beta (1-\beta
)\pi \alpha _{s}}{3m_{q}m_{Q}r^{2}}r^{-\beta }+V_{0};\text{ where }\beta
=1/2,3/4,  \label{51}
\end{equation}
respectively, which generate singlet states with opposite quark and
antiquark spins of the signature $n^{1}S_{0}.$ Furthermore, Eq. (49) can be
rewritten simply as

\[
V_{J=L}(r)=g(r)\left\{ \frac{1}{4}\frac{m_{Q}^{2}-m_{q}^{2}}{m_{q}m_{Q}}%
\sqrt{\frac{1}{10}(2L+3)(2L-1)}\left[ Ar^{-\beta }+(1-\epsilon )\kappa
r^{\beta }\right] \right. 
\]
\begin{equation}
-\left. \frac{1}{2}(1+\beta )(1-\epsilon )\kappa r^{\beta }\right\}
+V_{static}(r)  \label{52}
\end{equation}
which generates states of the signatures $%
n^{1}P_{1},~n^{1}D_{2},~n^{1}F_{3},~n^{1}G_{4},\cdots .$\ \ \ \ \ \ \ \ \ \
\ \ \ \ \ \ \ \ \ \ \ \ \ \ \ \ \ \ \ \ \ \ \ \ \ \ \ \ \ \ \ \ \ \ \ \ \ \
\ \ \ \ \ \ \ \ \ \ \ \ \ \ \ \ \ \ \ \ \ \ \ \ \ \ \ \ \ \ \ \ \ \ \ \ \ \
\ \ \ \ \ \ \ \ \ \ \ \ \ \ \ \ \ \ \ \ \ \ \ \ \ \ \ \ \ \ \ \ \ \ \ \ \ \
\ \ \ \ \ \ \ \ \ \ \ \ \ \ \ \ \ \ \ \ \ \ \ \ \ \ \ \ \ \ \ \ \ \ \ \ \ \
\ \ \ \ \ \ \ \ \ \ \ \ \ \ \ \ \ \ \ \ \ \ \ \ \ \ \ \ \ \ \ \ \ \ \ \ \ \
\ \ \ \ \ \ \ \ \ \ \ \ \ \ \ \ \ \ \ \ \ \ \ \ \ \ \ \ \ \ \ \ \ \ \ \ \ \ 

\subsection{Triplet states}

For triplet $(S=1)$ case, we have the known inequality $\left| L-S\right|
\leq J\leq L+S$ that gives $J=L$ and $~L\pm 1:$ \ \ \ \ \ \ \ \ \ \ \ \ \ \
\ \ \ \ \ \ \ \ \ \ \ \ \ \ \ \ \ \ \ \ \ \ \ \ \ \ \ \ \ \ \ \ \ \ \ \ \ \
\ \ \ \ 

\subsubsection{States $J=L$}

Here, the parity $P=(-1)^{J+1}$and the charge conjugation $C=(-1)^{L+1}.$
The potential in Eq. (44) takes the following simple form

\begin{equation}
V_{eff}(r)=V_{static}(r)+\frac{1}{4}\left( F_{SS}+F_{T}-4F_{LS}\right) +%
\sqrt{\frac{1}{10}(2L+3)(2L-1)}F_{LS_{-}},  \label{53}
\end{equation}
which can be substituted in (8) together with $\overline{k}=N+2J-a$ therein$%
. $ Further, the potential (53) reads

\[
V_{J=L}(r)=-\frac{g(r)}{2}\left\{ \left[ \left( \frac{4-\beta }{3}+\frac{1}{2%
}\frac{m_{q}^{2}+m_{Q}^{2}}{m_{q}m_{Q}}\right) Ar^{-\beta }+\left( 1+\frac{1%
}{2}\frac{m_{q}^{2}+m_{Q}^{2}}{m_{q}m_{Q}}\right) (1-2\epsilon )\kappa
r^{\beta }+\epsilon \kappa r^{\beta }\right] \right. 
\]

\begin{equation}
\left. -\frac{1}{2}\frac{m_{Q}^{2}-m_{q}^{2}}{m_{q}m_{Q}}\sqrt{\frac{1}{10}%
(2L+3)(2L-1)}\left[ Ar^{-\beta }+(1-\epsilon )\kappa r^{\beta }\right]
\right\} +V_{static},  \label{54}
\end{equation}
which generates states like $n^{3}P_{1},~n^{3}D_{2},~n^{3}F_{3},~n^{3}G_{4},%
\cdots .$

\subsubsection{States $J=L\pm 1$}

We have the parity $P=(-1)^{J}$ and the charge conjugation $C=(-1)^{L+1}.$
The eigenfunction is a superposition of two components with orbital momentum 
$L=J+1$ and $L=J-1$ which have equal space parity

\begin{equation}
\psi _{S,J}(r)=\widehat{u}_{J-1}(r)Y_{J-1,1,J}^{m}(\theta ,\varphi )+%
\widehat{u}_{J+1}(r)Y_{J+1,1,J}^{m}(\theta ,\varphi ).  \label{55}
\end{equation}
The action of the tensor operator, $T,$ on the two components of the
wavefunction in Eq. (55) is 
\begin{equation}
Tu_{J\pm 1}Y_{J\pm 1,1,J}^{m}(\widehat{{\bf r}})=\mp \frac{1}{4(2J+1)}%
u_{J\pm 1}Y_{J\pm 1,1,J}^{m}(\widehat{{\bf r}})+\frac{1}{2}\frac{\sqrt{J(J+1)%
}}{2J+1}u_{J\mp 1}Y_{J\mp 1,1,J}^{m}(\widehat{{\bf r}}).  \label{56}
\end{equation}
Therefore, a set of equations are obtained 
\[
\left\{ -\frac{1}{2\mu }\frac{d^{2}}{dr^{2}}+\frac{\left[ \overline{k}%
-\left( 1-a\right) \right] \left[ \overline{k}-\left( 3-a\right) \right] }{%
8\mu r^{2}}+V_{static}(r)+\frac{1}{4}\left( F_{SS}-\frac{F_{T}}{\left(
2J+1\right) }\right) -E_{n,J+1}\right. 
\]

\[
\left. -\left( J+2\right) F_{LS}-\frac{1}{2m%
%TCIMACRO{\UNICODE[m]{0xb4}}%
%BeginExpansion
{\acute{}}%
%EndExpansion
}\left[ V_{static}(r)-\left( J+2\right) F_{LS}+\frac{1}{4}\left( F_{SS}-%
\frac{F_{T}}{\left( 2J+1\right) }\right) -E_{n,J+1}\right] ^{2}\right\} 
\]

\begin{equation}
\times \widehat{u}_{n,J+1}(r)-\frac{\sqrt{J(J+1)}}{2\left( 2J+1\right) }F_{T}%
\widehat{u}_{n,J-1}(r)=0,  \label{57}
\end{equation}
and

\[
\left\{ -\frac{1}{2\mu }\frac{d^{2}}{dr^{2}}+\frac{\left[ \overline{k}%
-\left( 1-a\right) \right] \left[ \overline{k}-\left( 3-a\right) \right] }{%
8\mu r^{2}}+V_{static}(r)+\frac{1}{4}\left( F_{SS}+\frac{F_{T}}{\left(
2J+1\right) }\right) -E_{n,J-1}\right. 
\]

\[
\left. +\left( J-1\right) F_{LS}-\frac{1}{2m%
%TCIMACRO{\UNICODE[m]{0xb4}}%
%BeginExpansion
{\acute{}}%
%EndExpansion
}\left[ V_{static}(r)+\left( J-1\right) F_{LS}+\frac{1}{4}\left( F_{SS}+%
\frac{F_{T}}{\left( 2J+1\right) }\right) -E_{n,J-1}\right] ^{2}\right\} 
\]

\begin{equation}
\times \widehat{u}_{n,J-1}(r)-\frac{\sqrt{J(J+1)}}{2\left( 2J+1\right) }F_{T}%
\widehat{u}_{n,J+1}(r)=0,  \label{58}
\end{equation}
where $\overline{k}=N+2J+2-a.$ Therefore, Eqs. (57) and (58) describe states
such as $%
n^{3}P_{2},~n^{3}D_{3},~n^{3}F_{2},~n^{3}H_{4},n^{3}P_{0},~n^{3}D_{1},\cdots
.$ Here we may consider numerically the system obtained and separate
equations by dropping out the mixed terms to see their effect on the
spectrum of the masses. Consequently one can rewrite (57) and (58) in the
following simplest forms

\[
V_{J=L-1}(r)=-g(r)\left\{ (L+1)\left[ \left( 1+\frac{1}{4}\frac{%
m_{q}^{2}+m_{Q}^{2}}{m_{q}m_{Q}}\right) \left[ Ar^{-\beta }+\left(
1-2\epsilon \right) \kappa r^{\beta }\right] +\epsilon \kappa r^{\beta }%
\right] \right. 
\]

\[
+\frac{1}{4}\frac{1}{(2L-1)}\left[ (2+\beta )Ar^{-\beta }+(2-\beta )\left(
1-\epsilon \right) \kappa r^{\beta }\right] 
\]
\begin{equation}
\left. +\frac{1}{4}\left[ \frac{(2+\beta )}{3}Ar^{-\beta }-\beta \left(
1-\epsilon \right) \kappa r^{\beta }\right] \right\} +V_{static}(r),
\label{59}
\end{equation}
for states $n^{3}P_{0},~n^{3}D_{1},~n^{3}F_{2},~n^{3}H_{4},\cdots $ and

\[
V_{J=L+1}(r)=g(r)\left\{ \left[ \left( 1+\frac{1}{4}\frac{m_{q}^{2}+m_{Q}^{2}%
}{m_{q}m_{Q}}\right) \left[ Ar^{-\beta }+\left( 1-2\epsilon \right) \kappa
r^{\beta }\right] +\epsilon \kappa r^{\beta }\right] L\right. 
\]

\[
+\frac{1}{4}\frac{1}{(2L+3)}\left[ (2+\beta )Ar^{-\beta }+(2-\beta )\left(
1-\epsilon \right) \kappa r^{\beta }\right] 
\]
\begin{equation}
\left. -\frac{1}{4}\left[ \frac{(2+\beta )}{3}Ar^{-\beta }-\beta \left(
1-\epsilon \right) \kappa r^{\beta }\right] \right\} +V_{static}(r),
\label{60}
\end{equation}
for states $n^{3}P_{2},~n^{3}D_{3},\cdots .$ Further, for triplet $S$-wave,
we have 
\begin{equation}
V_{eff}(r)=-\frac{A}{r}+\kappa r+\frac{8\pi \alpha _{s}}{9m_{q}m_{Q}}\delta
^{(3)}({\bf r})+V_{0},  \label{61}
\end{equation}
and 
\begin{equation}
V_{eff}(r)=-\frac{A}{r^{\beta }}+\kappa r^{\beta }+\frac{2\beta (1-\beta
)\pi \alpha _{s}}{9m_{q}m_{Q}r^{2}}r^{-\beta }+V_{0};\text{ where }\beta
=1/2,3/4,  \label{62}
\end{equation}
which describe states such as $n^{3}S_{1.}$

\subsubsection{State $J=0$}

Equations (59) and (60) degenerate into a single equation with an effective
potential

\begin{equation}
V_{eff}(r)=V_{static}(r)+\frac{1}{4}(F_{SS}-F_{T}-8F_{LS}),  \label{63}
\end{equation}
and also by setting $\overline{k}=N+2-a$ therein$.$ Further, Eq. (63) becomes

\[
V_{J=0}(r)=-\left\{ 1+g(r)\left[ 2+\frac{1}{2}\frac{m_{q}^{2}+m_{Q}^{2}}{%
m_{q}m_{Q}}+\frac{2+\beta }{3}\right] \right\} Ar^{-\beta } 
\]

\begin{equation}
+\left\{ 1-\frac{g(r)}{2}\left[ \left( 5-\beta \right) \left( 1-\epsilon
\right) +\left( \frac{m_{q}^{2}+m_{Q}^{2}}{m_{q}m_{Q}}\right) \left(
1-2\epsilon \right) \right] \right\} \kappa r^{\beta }+V_{0}.  \label{64}
\end{equation}
which only describes states such as $n^{3}P_{0}.$

\section{PSEUDOSCALAR AND VECTOR DECAY CONSTANTS OF THE $B_{c}$ MESON}

The significant contribution to the $B_{c} $ total decay rate comes from the
annihilation of the $c$ quark and $\overline{b}$ antiquark into the vector
boson $W^{+}$ which decays into a lepton and a neutrino or a quark-antiquark
pair.

The nonrelativistic expression for the decay constants is given by [45-47]

\begin{equation}
f_{P}^{NR}=f_{V}^{NR}=\sqrt{\frac{12}{M_{P,V}(q\overline{Q})}}\left| \Psi
_{P,V}(0)\right|  \label{65}
\end{equation}
where $\Psi _{P,V}(0)$ is the meson wave function at the origin $r=0.$ The $%
f_{P}$ and $f_{V}$, $P$ corresponds to the pseudoscalar $B_{c}$ and $V$ \ to
to the vector $B_{c}^{\ast }$ mesons and $M_{P,V}(q\overline{Q})$ are the
masses of the $B_{c}$ and $B_{c}^{\ast }$ mesons.

\section{RESULTS AND CONCLUSION}

We have solved the spinless Salpeter equation using SLNET and also extended
our earlier formalism for the SAD spectra [31] by introducing the spin
corrections. Furthermore, we have obtained a unified description of the
self-conjugate meson spectroscopy with a phenomenological and a
QCD-motivated potential model. This model has also extended to comprise
various cases of pure scalar confinement $(\epsilon =1),$ scalar-vector
couplings $(\epsilon =1/2)$ and the vector confinement $(\epsilon =0)$
interactions. It was motivated by our desire to construct an analytic
expression for the quarkonium mass spectra of different spins and confining
interactions that could have combined both heavy and light quarkonia.

The parameter fits used in this work together with the quark masses are
shown in Table I. They are taken to be same as those considered in our
earlier works [10] for each static attractive potential. We do not view our
present result as a totally successful model of quarkonia interactions even
though the level structure is good and the transitions are also well
accounted for. It seems likely that in this study our model has shown
remarkable success in describing the heavy quarkonia data well once the
potential parameters are fitted properly for optimum agreement with the
experimentally observed data.

We have made the Salpeter binding mass Eq. (26) more general by including
the spin-dependent corrections to the potential. Such a formula is able to
describe with good accuracy the spectra of all quark-antiquark bound states
as long as the mass of the constituent quark is larger than the term $%
W_{n,l}(r)$ as remarked earlier. Our predictions for the Salpeter mass
spectrum of all unobserved and observed charmonium and upsilon systems, to
all static potentials, in the flavour-dependent and flavour-independent
cases with $\epsilon =0,1/2,1$ are presented in Tables II-III. Further,
Table IV shows the predicted Salpeter masses of unobserved $n^{2S+1}D_{J},$ $%
n^{2S+1}F_{J}$ and $n^{2S+1}G_{J}$ bottomonium levels using a new set of
potential parameters. Therefore the calculated bottom masses of all static
potentials for the $n^{2S+1}S_{J},$ $n^{2S+1}P_{J}$ and $n^{2S+1}D_{J}$
states are quite fair for the Turin potential as clearly seen from the
average relative error range, $\left\langle
(M^{Th}-M^{Exp})/M^{Exp}\right\rangle $. We have also predicted the $B_{c}$
meson spectrum for the three static potentials in Table V using the set of
parameters in Table I. In Tables VI and VII we calculated the $c\overline{b}%
, $ $c\overline{c}$ and $b\overline{b}$ using a different set of fitted
parameters for Cornell potential. No additive constant is permitted in this
manner and the accuracy in producing spectra is pretty good. There is a
clear preference in these fits for the Salpeter wave equation with an
approximately equal mixture of scalar and vector $(\epsilon =1/2)$ couplings
and also vector confinement $(\epsilon =0).$ We found that larger values of $%
\epsilon $ are mildly preferred by fits to the spin-singlet or triplet
states while the quality of the $\epsilon =1$ fit to the spin singlet and
triplet data is not as high as that of the $\epsilon =1/2$ and $\ \epsilon
=0 $ fits, it is certainly acceptable, with a value of errors. We conclude
here that the Lorentz structure of the confining interaction cannot be
determined using only $\epsilon =1/2.$ We will henceforth restrict our
attention to the case of pure scalar confinement at large distance as
expected theoretically. The three potentials seem to be fairly good in
fitting all the data..

In the equal scalar and vector couplings, we have found that our fits are
very good with level values and accurate to a few ${\rm MeV}.$ For
convenience we compare explicitly the predicted and measured spin splitting
energy for different $L$ states. We find that the apparent success is
achieved for the predicted $\chi _{b2}-\chi _{b1}=24~{\rm MeV}$ and $\chi
_{b1}-\chi _{b0}=33~{\rm MeV}$ in the average for the three potentials and
are very close to the experimental values $21~{\rm MeV}$ and 32$~{\rm MeV}$
respectively. Furthermore, the predicted $\chi _{b2}^{\prime }-\chi
_{b1}^{\prime }=13~{\rm MeV}$ and $\chi _{b1}^{\prime }-\chi _{b0}^{\prime
}=16~{\rm MeV}$ in the average for the three potentials which are exactly
same as the experimental value $13~{\rm MeV}$ and close to 2$3~{\rm MeV},$
respectively. The predicted hyperfine splitting $\Delta _{{\rm hfs}}{\rm (1S)%
}=M(\Upsilon {\rm (1S)})-M(\eta _{b}{\rm (1S)})=80_{-8}^{+6}~{\rm MeV,}$
(cf. [33])$,$ $\Delta _{{\rm hfs}}{\rm (2S)}=M(\Upsilon ^{\prime }{\rm (2S)}%
)-M(\eta _{b}^{\prime }{\rm (2S)})=22_{-2}^{+3}~{\rm MeV,}$ and $\Delta _{%
{\rm hfs}}{\rm (3S)}=M(\Upsilon ^{\prime \prime }{\rm (3S)})-M(\eta
_{b}^{\prime \prime }{\rm (3S)})=14_{-1}^{+2}~{\rm MeV}$ are nearly close to
the theoretically calculated values 62$~{\rm MeV},$ $40$ ${\rm MeV,}$ and $15
$ ${\rm MeV,}$ respectively, (cf. [33,39]). Further, The hyperfine splitting
for the ${\rm 2S}$ charmonium state is calculated and the predicted number
is $\Delta _{{\rm hfs}}{\rm (2S)}=M(\psi {\rm (2S)})-M(\eta _{c}{\rm (2S)}%
)=56_{-8}^{+4}~{\rm MeV}$ for flavour dependent case and $56_{-8}^{+18}~{\rm %
MeV}$ for flavour independent case, (cf. [37,39]) . Badalian and Bakker in
their recent work calculated and predicted the number as $\Delta _{{\rm hfs}}%
{\rm (2S,}${\rm theory}${\rm )=}57\pm 8$ ${\rm MeV}$ [37] giving $M(\eta _{c}%
{\rm (2S})=$3630$\pm 8$ ${\rm MeV.}$

We observe, in this connection, that splitting approximation can be improved
significantly by increasing the quantum number $L.$ It has also been noted
that the term contributing most to the Salpeter binding energy is the
leading term $E_{0}$ as $L$ becomes non zero which improves the convergence
of Eq. (25). This is expected since the expansion parameter $1/\overline{k}$
becomes smaller as $L$ becomes larger.

It is found that most results agree moderately well with the experimental
data, or with the theoretical predictions wherever experimental data are not
available. Our predictions for the lower states are considerably better than
the higher states since $\Upsilon (n{\rm S}),$ where $n>4$ lie above the
threshold and predominantly decay into mesons containing charm and beauty
flavour.\footnote{%
In the upsilon family, Kiselev [48] found $n=4.$}

The results obtained are a relevant summary of the effectiveness of the
method. It seems likely, that the average relative errors of the fitting
results are almost very small for most states. The comparisons with the
experimental data show clearly that when we are taking into account the
relativistic corrections and also the third order approximation to the
binding energy, the coincidence with the experimental data is improved. The
deviations from experiment are more considerable at $n^{1}{\rm S}_{0}$
states for c$\overline{c}$ and $b\overline{b}$ systems. These results show
that the last terms in these equations in most cases are not important for
the spectrum of meson masses. We do not pretend to display a whole account
of the meson spectroscopy. Our purpose is to demonstrate the possibility of
applicance of SLNET for determination of meson masses in the context of the
SS equation using relativistic kinematics.

We have found that fitting the parameters is extremely essential to enable
one to sharpen the analysis.\footnote{%
We have used the available parameter fits in the literature as a test for
the present work.} The prescription and the wave equation used are
responsible for the deterioration of the fit parameters. The calculation and
parameters are also model dependent as remarked in our earlier papers
[10,31]. All we can say that the SLNET works for the SS equation well with
relativistic kinematics as these parameters are fitted properly and once the
mass of the quark is taken large.

>From a phenomenological point of view, a certain amount of flavour
dependence is required. It was pointed out by Miller {\it et al. }in [14]
that no flavour-independent potential could fit both the spin-averaged c$%
\overline{c}$ and $b\overline{b}$ levels.

It is clear that the coulomb-like parameter $A$ is in accordance with the
ideas of asymptotic freedom is expected for the strong gauge-coupling
constant of QCD, that is, $\alpha _{s}(m_{c})\neq \alpha _{s}(m_{b})\neq
\alpha _{s}(\mu )$ with $\mu =m_{c}m_{b}/(m_{c}+m_{b});$ (cf. e.g., Ref.
[48])$.$\footnote{%
For a pretty good fit to the $c\overline{c},$ $b\overline{b}$ and $c%
\overline{b}$ quarkonium spectra, the QCD coupling constant $\alpha _{s}(\mu
^{2})$ must be dependent on the quark-flavour. Kiselev [48] took $\alpha
_{s}(m_{b})/\alpha _{s}(m_{c})\simeq 3/4.$} Of course a complete analysis
would require ingredients that we have not considered here but discussed in
the literature.

The calculated values of the pseudoscalar and vector decay constants of the $%
B_{c}$ meson in our model using the nonrelativistic formula (65) are
displayed in Table VIII. The radial wave function at the origin is being
calculated in Table IX. They are compared with the ones calculated using the
relativistic and other predictions using the nonrelativistic quark models.
The calculated values of these leptonic constants are consistent with the
other predictions [22,25,29,38,49,50]. Kiselev {\it et al}. [49] estimated
the pseudoscalar leptonic constant $f_{B_{c}}^{NR}=493~{\rm MeV}$ in the
potential model. Furthermore, the calculations in the same potential model
with the one-loop matching [50] is $f_{B_{c}}^{1-loop}=400\pm 45~{\rm MeV}$
and also for the two-loop calculations [50] is $f_{B_{c}}^{2-loop}=395\pm 15~%
{\rm MeV}.$

We may expect to improve our results by adding extra spin-independent terms
comparable to the ones induced by the original scalar confinement potential.

\acknowledgments
(S. M. Ikhdair) is grateful to his wife, Oyoun, and his son, Musbah, for
their love, understanding, and patience. Their encouragement provided the
necessary motivation to complete this work.\newpage

\appendix

\section{The Spin-Correction Terms:}

For parastates $(S=0)$ case we have:

\begin{equation}
J=L
\end{equation}
For triplet $(S=1)$ case we have the following:

\begin{equation}
J=\left\{ 
\begin{array}{l}
L-1,\text{ }{\bf S}\cdot {\bf L}=-(L+1)~ \\ 
L,~{\bf S}\cdot {\bf L}=-1 \\ 
L+1,\text{ }{\bf S}\cdot {\bf L}=L
\end{array}
\right.
\end{equation}
For bound-state constituents of spin $S_{1}=S_{2}=1/2,$ the independent
operators ${\bf S}_{1}\cdot {\bf S}_{2},$:$\left( {\bf S}_{1}\pm {\bf S}%
_{2}\right) \cdot {\bf L}$ and $T:$

\begin{equation}
\left\langle {\bf S}_{1}\cdot {\bf S}_{2}\right\rangle =\left\{ 
\begin{array}{l}
-3/4\text{ for spin singlets, }S=0,\text{ } \\ 
+1/4\text{ for spin triplets, }S=1.
\end{array}
\right.
\end{equation}

\begin{equation}
\left\langle {\bf S}\cdot {\bf L}\right\rangle =\left\{ 
\begin{array}{l}
0,\text{ for spin singlets }S=0,\text{ } \\ 
\frac{1}{2}\left[ J(J+1)-L(L+1)-2\right] ,\text{ for spin triplets }S=1.
\end{array}
\right.
\end{equation}

\begin{equation}
({\bf S}_{1}\cdot \widehat{{\bf r}}{\bf S}_{2}\cdot \widehat{{\bf r}}%
)u_{J}(r)Y_{J,0,J}(\widehat{{\bf r}})=-\frac{1}{4}u_{J}(r)Y_{J,0,J}(\widehat{%
{\bf r}}),
\end{equation}

\begin{equation}
({\bf S}_{1}\cdot {\bf S}_{2})Y_{J,S,L}^{m}(\widehat{{\bf r}})=\frac{1}{2}%
\left[ S(S+1)-S_{1}(S_{1}+1)-S_{2}(S_{2}+1)\right] Y_{J,S,L}^{m}(\widehat{%
{\bf r}}),
\end{equation}

\begin{equation}
\left( {\bf S}_{1}+{\bf S}_{2}\right) \cdot {\bf L}Y_{J,S,L}^{m}(\widehat{%
{\bf r}}){\bf =}\frac{1}{2}\left[ J(J+1)-L(L+1)-S(S+1)\right] Y_{J,S,L}^{m}(%
\widehat{{\bf r}}),
\end{equation}

\begin{equation}
\left( {\bf S}_{1}-{\bf S}_{2}\right) \cdot {\bf L}Y_{J,S,L}^{m}(\widehat{%
{\bf r}}){\bf =}\sqrt{\frac{1}{10}\left[ 2L+3)(2L-1)\right] }\delta
_{J,L}\left( \delta _{S,0}Y_{J,1,L}^{m}(\widehat{{\bf r}})+\delta
_{S,1}Y_{J,0,L}^{m}(\widehat{{\bf r}})\right) ,
\end{equation}

\[
TY_{J,1,L}^{m}(\widehat{{\bf r}}){\bf =}\frac{1}{4}\delta
_{J,L}Y_{J,1,L}^{m}(\widehat{{\bf r}})-\frac{1}{4(2L-1)}\delta
_{J,L-1}Y_{J,1,L}^{m}(\widehat{{\bf r}})+\frac{1}{4(2L+3)}\delta
_{J,L+1}Y_{J,1,L}^{m}(\widehat{{\bf r}}) 
\]

\begin{equation}
-\frac{\sqrt{(L+1)(L+2)}}{2(2L+3)}\delta _{J,L+1}Y_{J,1,L+2}^{m}(\widehat{%
{\bf r}})-\frac{\sqrt{L(L-1)}}{2(2L-1)}\delta _{J,L-1}Y_{J,1,L-2}^{m}(%
\widehat{{\bf r}}).
\end{equation}

\bigskip

\newpage

\bigskip \baselineskip= 2\baselineskip% double space the text

\bigskip 
\begin{table}[tbp]
\caption{Fitted parameters of the class of static central potentials. }
\label{table1}
\begin{tabular}{llll}
& Cornell & Song-Lin & Turin \\ 
\tableline$c\overline{c}$ &  &  &  \\ 
$m_{c}~$ & 1.84$~GeV$ & 1.80$~GeV$ & 1.82$~GeV$ \\ 
$A$ & 0.509 & 0.498$~GeV^{1/2}$ & 0.547$~GeV^{1/4}$ \\ 
$\kappa ~$ & 0.166$~GeV^{2}$ & 0.694$~GeV^{3/2}$ & 0.309$~GeV^{7/4}$ \\ 
$V_{0}~$ & -0.799$~GeV$ & -1.314$~GeV$ & -0.898$~GeV$ \\ 
\tableline$b\overline{b}$ &  &  &  \\ 
$m_{b}~$ & 5.17$~GeV$ & 5.20$~GeV$ & 5.18$~GeV$ \\ 
$A$ & 0.443 & 0.762$~GeV^{1/2}$ & 0.576$~GeV^{1/4}$ \\ 
$\kappa ~$ & 0.184$~GeV^{2}$ & 0.576$~GeV^{3/2}$ & 0.304$~GeV^{7/4}$ \\ 
$V_{0}~$ & -0.805$~GeV$ & -0.991$~GeV$ & -0.845$~GeV$ \\ 
\tableline Combined &  &  &  \\ 
$m_{c}~$ & 1.79$~GeV$ & 1.82$~GeV$ & 1.80$~GeV$ \\ 
$m_{b}~$ & 5.17$~GeV$ & 5.20$~GeV$ & 5.18$~GeV$ \\ 
$A$ & 0.457 & 0.722$~GeV^{1/2}$ & 0.578$~GeV^{1/4}$ \\ 
$\kappa ~$ & 0.182$~GeV^{2}$ & 0.589$~GeV^{3/2}$ & 0.304$~GeV^{7/4}$ \\ 
$V_{0}~$ & -0.790$~GeV$ & -1.029$~GeV$ & -0.838$~GeV$%
\end{tabular}
\end{table}

\bigskip

\widetext

\bigskip 
\begin{table}[tbp]
\caption{Salpeter masses of the unobserved $c\overline{c}$ and $b\overline{b}
$ for the flavor-dependent and flavor-independent cases, respectively for
the Cornell, Song-Lin and Turin potentials}
\begin{tabular}{llllllll}
State\tablenotemark[1] & [45] & Cornell\tablenotemark[2] & Song-Lin%
\tablenotemark[2] & Turin\tablenotemark[2] & Cornell\tablenotemark[3] & 
Song-Lin\tablenotemark[3] & Turin\tablenotemark[3] \\ 
\tableline$\eta _{c}(2S)$\tablenotemark[4] & 3589 & 3604 & 3595 & 3593 & 3602
& 3574 & 3585 \\ 
(1$^{1}D_{2})c\overline{c}$ & 3826 & 3755 & 3756 & 3756 & 3748 & 3756 & 3754
\\ 
(1$^{3}D_{1})c\overline{c}$ & 3814 & 3708 & 3708 & 3705 & 3699 & 3703 & 3701
\\ 
$\eta _{b}(1S)$ & 9314 & 9398 & 9407 & 9395 & 9380 & 9433 & 9399 \\ 
$\eta _{b}(2S)$ & 9931 & 10000 & 10000 & 9997 & 10000 & 10014 & 10003 \\ 
$\eta _{b}(3S)$ & 10288 & 10342 & 10329 & 10332 & 10343 & 10341 & 10338 \\ 
$\eta _{b}(4S)$ & 10577 & 10612 & 10570 & 10586 & 10613 & 10581 & 10592 \\ 
$(1^{1}P_{1})b\overline{b}$ & 9906 & 9905 & 9890 & 9897 & 9908 & 9901 & 9902
\\ 
$(2^{1}P_{1})b\overline{b}$ & 10254 & 10246 & 10246 & 10245 & 10249 & 10256
& 10251 \\ 
$(1^{1}D_{2})b\overline{b}$\tablenotetext[1]{States with equal mixtures of
scalar and vector couplings.} & 10155 & 10142 & 10149 & 10146 & 10145 & 10157
& 10152 \\ 
$(2^{1}D_{2})b\overline{b}$\tablenotetext[2]{Here's the flavour-dependent
case.} & 10450 & 10424 & 10422 & 10423 & 10427 & 10431 & 10429 \\ 
$\chi _{b0}(3P)$\tablenotetext[3]{Here's the flavour-independent case.} & 
10519 & 10512 & 10491 & 10498 & 10513 & 10501 & 10504 \\ 
$\chi _{b1}(3P)$\tablenotetext[4]{Recently observed by Belle Collaboration
[34,35].} & 10542 & 10520 & 10500 & 10508 & 10522 & 10510 & 10514 \\ 
$\chi _{b2}(3P)$ & 10561 & 10528 & 10508 & 10516 & 10530 & 10518 & 10522 \\ 
$(1^{3}D_{1})b\overline{b}$ & 10147 & 10128 & 10132 & 10131 & 10131 & 10142
& 10137 \\ 
$(1^{3}D_{2})b\overline{b}$ & 10153 & 10140 & 10147 & 10145 & 10144 & 10156
& 10151 \\ 
$(1^{3}D_{3})b\overline{b}$ & 10158 & 10152 & 10161 & 10158 & 10156 & 10170
& 10164 \\ 
$(2^{3}D_{1})b\overline{b}$ & 10442 & 10414 & 10411 & 10412 & 10417 & 10420
& 10418 \\ 
$(2^{3}D_{2})b\overline{b}$ & 10448 & 10423 & 10421 & 10422 & 10426 & 10430
& 10428 \\ 
$(2^{3}D_{3})b\overline{b}$ & 10453 & 10433 & 10431 & 10432 & 10436 & 10440
& 10438
\end{tabular}
\end{table}

\bigskip \widetext

\bigskip

\begin{table}[tbp]
\caption{Salpeter $c\overline{c}$ and $b\overline{b}$ quarkonium masses of
the observed states (in $MeV)$ for the Cornell, Song-Lin and Turin
potentials }
\label{table3}
\begin{tabular}{llllllllll}
State\tablenotemark[1] & $M^{(\exp )}[36]$ & [45] & Cornell\tablenotemark[2]
& Song-Lin\tablenotemark[2] & Turin\tablenotemark[2] & Cornell%
\tablenotemark[3] & Song-Lin\tablenotemark[3] & Turin\tablenotemark[3] &  \\ 
\tableline$\eta _{c}(1S)$ & 2980 & 2967 & 2981 & 2965 & 2950 & 2988 & 2983%
\tablenotemark[4] & 2999\tablenotemark[4] &  \\ 
$\psi (1S)$ & 3097 & 3126 & 3104 & 3108 & 3111 & 3092 & 3096\tablenotemark[4]
& 3087\tablenotemark[4] &  \\ 
$\psi (2S)$ & 3686 & 3700 & 3652 & 3655 & 3652 & 3650 & 3640 & 3639 &  \\ 
$\chi _{c0}(1P)$ & 3415 & 3427 & 3316 & 3374 & 3335 & 3322 & 3340 & 3309 & 
\\ 
$\chi _{c1}(1P)$ & 3511 & 3497 & 3471 & 3461 & 3464 & 3456 & 3463 & 3459 & 
\\ 
$\chi _{c2}(1P)$ & 3556 & 3543 & 3535 & 3521 & 3530 & 3521 & 3532 & 3529 & 
\\ 
$h_{c}(1P)$ & 3526 & 3510 & 3488 & 3468 & 3478 & 3471 & 3475 & 3474 &  \\ 
$\Upsilon (1S)$ & 9460 & 9511 & 9474 & 9499 & 9488 & 9466 & 9505%
\tablenotemark[4] & 9480\tablenotemark[4] &  \\ 
$\Upsilon (2S)$ & 10023 & 10004 & 10019 & 10034 & 10024 & 10020 & 10039%
\tablenotemark[4] & 10025\tablenotemark[4] &  \\ 
$\Upsilon (3S)$ & 10355 & 10350 & 10354 & 10350 & 10349 & 10356 & 10357 & 
10352 &  \\ 
$\Upsilon (4S)$ & 10580 & 10642 & 10622 & 10585 & 10599 & 10624 & 10593 & 
10603 &  \\ 
$\chi _{b0}(1P)$ & 9860 & 9854 & 9863 & 9854 & 9855 & 9863 & 9868 & 9860 & 
\\ 
$\chi _{b1}(1P)$ & 9892 & 9878 & 9898 & 9885 & 9890 & 9900 & 9885 & 9896 & 
\\ 
$\chi _{b2}(1P)$ & 9913 & 9896 & 9921 & 9910 & 9915 & 9924 & 9920 & 9921 & 
\\ 
$\chi _{b0}(2P)$ & 10232 & 10213 & 10228 & 10228 & 10226 & 10230 & 10239 & 
10232 &  \\ 
$\chi _{b1}(2P)$ & 10255 & 10236 & 10243 & 10244 & 10242 & 10245 & 10254 & 
10248 &  \\ 
$\chi _{b2}(2P)$ & 10268 & 10254 & 10255 & 10257 & 10256 & 10258 & 10266 & 
10262 & 
\end{tabular}
\tablenotetext[1]{States with equal mixtures of scalar and vector couplings.}%
\tablenotetext[2]{Here's the flavour-dependent case.}%
\tablenotetext[3]{Here's the flavour-independent case.}%
\tablenotetext[4]{The
hyperfine splitting term is not considered here.}
\end{table}

\bigskip

\bigskip \mediumtext

\begin{table}[tbp]
\caption{Salpeter $b\overline{b}$ quarkonium masses (in $MeV)$ for the
Cornell, Song-Lin and Turin potentials}
\label{table5}
\begin{tabular}{llllllll}
State\tablenotemark[1] & [45] & Cornell & Error \% & Song-Lin & Error \% & 
Turin & Error \% \\ 
\tableline$1^{1}S_{0}$\tablenotemark[2] & 9398 & 9342 & 0.60 & 9344 & 0.57 & 
9343 & 0.58 \\ 
$1^{3}S_{1}$ & 9460 & 9438 & 0.23 & 9464 & 0.04 & 9457 & 0.03 \\ 
$1^{3}P_{2}$ & 9913 & 9912 & 0.01 & 9900 & 0.13 & 9907 & 0.06 \\ 
$1^{3}P_{1}$ & 9891 & 9886 & 0.05 & 9872 & 0.19 & 9879 & 0.12 \\ 
$1^{3}P_{0}$ & 9870 & 9846 & 0.24 & 9834 & 0.36 & 9838 & 0.32 \\ 
$1^{1}P_{1}$ & 9908 & 9894 & 0.14 & 9878 & 0.30 & 9887 & 0.21 \\ 
$1^{3}D_{3}$ & - & 10149 &  & 10162 &  & 10157 &  \\ 
$1^{3}F_{4}$ & - & 10335 &  & 10356 &  & 10348 &  \\ 
$1^{3}G_{5}$ & - & 10496 &  & 10514 &  & 10507 &  \\ 
2$^{1}S_{0}$ & 9983 & 9986 & 0.03 & 9983 & 0.0 & 9984 & 0.01 \\ 
2$^{3}S_{1}$ & 10023 & 10008 & 0.15 & 10024 & 0.01 & 10015 & 0.08 \\ 
3$^{3}S_{1}$ & 10355 & 10351 & 0.04 & 10349 & 0.06 & 10347 & 0.08 \\ 
4$^{3}S_{1}$ & 10580 & 10623 & 0.41 & 10588 & 0.07 & 10601 & 0.20 \\ 
Fits\tablenotemark[3] & $A$ & $0.470$ &  & $0.869~GeV^{1/2}$ &  & $%
0.620~GeV^{1/4}$ &  \\ 
& $\kappa $ & $0.186~GeV^{2}$ &  & $0.558~GeV^{3/2}$ &  & $0.304~GeV^{7/4}$
&  \\ 
& $V_{0}$ & $-0.802~GeV$ &  & $-0.893~GeV$ &  & $-0.823~GeV$ &  \\ 
& $m_{b}$ & $5.17~GeV$ &  & $5.20~GeV$ &  & $5.18~GeV$ & 
\end{tabular}
\tablenotetext[1]{States with equal mixtures of scalar and vector couplings.}%
\tablenotetext[2]{Recently searched by ALEPH Collaboration [33].}%
\tablenotetext[3]{Here we cite Ref. [10] for the parameter fits.}
\end{table}

\bigskip

\bigskip

\bigskip \widetext
\begin{table}[tbp]
\caption{$B_{c}$ meson mass spectrum (in $MeV$) for the Cornell, Song-Lin
and Turin potentials, with a flavour-independent case.}
\label{table10}
\begin{tabular}{llllllllllll}
State & $\epsilon =1$ & $\epsilon =1/2$ & $\epsilon =0$ & $\epsilon =1$ & $%
\epsilon =1/2$ & $\epsilon =0$ & $\epsilon =1$ & $\epsilon =1/2$ & $\epsilon
=0$ & [45] & [50] \\ 
\tableline1$^{1}S_{0}$ & 6296.9 & 6284.9 & 6272.7 & 6287.2 & 6269.3 & 6250.2
& 6278.6 & 6263.3 & 6247.3 & 6270 & 6253 \\ 
2$^{1}S_{0}$ & 6852.5 & 66846.7 & 6840.8 & 6855.0 & 6848.5 & 6841.9 & 6850.0
& 6843.9 & 6837.7 & 6835 & 6867 \\ 
3$^{1}S_{0}$ & 7220.1 & 7215.9 & 7211.7 & 7190.4 & 7186.7 & 7182.8 & 7200.9
& 7196.9 & 7193.0 & 7193 &  \\ 
4$^{1}S_{0}$ & 7520.3 & 7516.9 & 7513.5 & 7441.1 & 7438.5 & 7435.9 & 7474.4
& 7471.4 & 7468.5 &  &  \\ 
$1^{1}P_{1}$ & 6724.8 & 6719.9 & 6714.9 & 6726.6 & 6721.8 & 6717.0 & 6725.9
& 6721.1 & 6716.2 & 6734 & 6717 \\ 
2$^{1}P_{1}$ & 7100.4 & 7096.6 & 7092.7 & 7091.7 & 7088.5 & 7085.3 & 7095.0
& 7091.5 & 7088.1 & 7126 & 7113 \\ 
$1^{1}D_{2}$ & 6979.9 & 6976.5 & 6973.0 & 6992.9 & 6990.3 & 6987.7 & 6988.3
& 6985.3 & 6982.3 & 7077 & 7001 \\ 
2$^{1}D_{2}$ & 7298.9 & 7296.0 & 7293.0 & 7277.4 & 7275.4 & 7273.4 & 7287.4
& 7285.0 & 7282.5 &  &  \\ 
$1^{3}P_{1}$ & 6717.5 & 6706.5 & 6695.4 & 6724.9 & 6710.9 & 6696.5 & 6719.7
& 6707.5 & 6695.0 & 6749 & 6729 \\ 
2$^{3}P_{1}$ & 7098.4 & 7090.1 & 7081.8 & 7091.7 & 7082.7 & 7073.4 & 7093.1
& 7084.6 & 7076.0 & 7145 & 7124 \\ 
$1^{3}D_{2}$ & 6981.4 & 6974.0 & 6966.6 & 6994.5 & 6987.2 & 6979.8 & 6989.4
& 6982.2 & 6974.9 & 7079 & 7016 \\ 
2$^{3}D_{2}$ & 7300.4 & 7294.0 & 7287.7 & 7278.7 & 7273.1 & 7267.4 & 7288.4
& 7282.6 & 7276.6 &  &  \\ 
1$^{3}S_{1}$ & 6296.9 & 6300.8 & 6304.8 & 6322.7 & 6327.6 & 6332.4 & 6310.1
& 6314.4 & 6318.6 & 6332 & 6317 \\ 
2$^{3}S_{1}$ & 6852.5 & 6854.4 & 6856.4 & 6862.4 & 6864.5 & 6866.6 & 6854.4
& 6856.3 & 6858.3 & 6881 & 6902 \\ 
3$^{3}S_{1}$ & 7220.1 & 7221.5 & 7222.9 & 7193.5 & 7194.7 & 7196.0 & 7202.5
& 7203.8 & 7205.1 & 7235 &  \\ 
4$^{3}S_{1}$ & 7520.3 & 7521.4 & 7522.6 & 7442.8 & 7443.6 & 7444.5 & 7475.2
& 7476.2 & 7477.2 &  &  \\ 
$1^{3}P_{2}$ & 6735.9 & 6748.5 & 6762.0 & 6739.2 & 6755.1 & 6771.8 & 6747.0
& 6753.6 & 6768.0 & 6762 & 6743 \\ 
2$^{3}P_{2}$ & 7104.7 & 7115.6 & 7126.4 & 7097.9 & 7109.2 & 7120.2 & 7101.4
& 7112.2 & 7122.8 & 7156 & 7134 \\ 
$1^{3}D_{3}$ & 6979.4 & 6998.1 & 7016.3 & 6996.1 & 7013.6 & 7030.9 & 6990.4
& 7008.3 & 7025.3 & 7081 & 7007 \\ 
2$^{3}D_{3}$ & 7298.3 & 7314.4 & 7331.1 & 7279.4 & 7293.3 & 7307.5 & 7288.9
& 7303.6 & 7318.8 &  &  \\ 
$1^{3}P_{0}$ & 6673.1 & 6639.2 & 6602.9 & 6695.8 & 6652.0 & 6602.0 & 6676.7
& 6636.7 & 6592.3 & 6699 & 6683 \\ 
2$^{3}P_{0}$ & 7082.4 & 7060.7 & 7038.6 & 7078.0 & 7052.4. & 7025.2 & 7075.9
& 7052.5 & 7028.2 & 7091 & 7088 \\ 
$1^{3}D_{1}$ & 6977.4 & 6949.6 & 6921.2 & 6988.9 & 6960.1 & 6929.5 & 6983.2
& 6955.1 & 6926.0 & 7072 & 7008 \\ 
2$^{3}D_{1}$ & 7297.8 & 7274.1 & 7250.2 & 7274.8 & 7252.9 & 7230.0 & 7284.1
& 7261.6 & 7238.5 &  & 
\end{tabular}
\end{table}

\bigskip

\bigskip

\widetext
\begin{table}[tbp]
\caption{Flavor-independent $c\overline{b}$ mass spectrum (in $MeV$) using
the Cornell potential with scalar ($\protect\epsilon =1),$ an equal mixture
of scalar-vector ($\protect\epsilon =1/2)$ and vector ($\protect\epsilon =0)$
interactions. }
\label{table11}
\begin{tabular}{lllllllll}
State & $\epsilon =1$\tablenote{Here $\epsilon=1$ is the scalar confinement
interaction.} & $\epsilon =1/2$\tablenote{Here $\epsilon=1/2$ is the equal
mixture of scalar-vector confinement interaction.} & $\epsilon =0$%
\tablenote{Here $\epsilon=0$ is the vector confinement interaction.} & State
& $\epsilon =1$ & $\epsilon =1/2$ & $\epsilon =0$ & Parameters [25] \\ 
\tableline$1^{1}S_{0}$ & 6302.6 & 6284.5 & 6265.9 & 1$^{3}S_{1}$ & 6302.6 & 
6308.5 & 6314.4 & $A=0.437$ \\ 
2$^{1}S_{0}$ & 6884.6 & 6875.2 & 6865.6 & 2$^{3}S_{1}$ & 6884.6 & 6887.7 & 
6890.8 & $\kappa =0.203~GeV^{2}$ \\ 
3$^{1}S_{0}$ & 7285.1 & 7278.2 & 7271.2 & 3$^{3}S_{1}$ & 7285.1 & 7287.4 & 
7289.7 & $V_{0}=0$\tablenote{It's a shifting parameter (in $MeV$) to be set
to fix the the position of $M(B_{c})$ and $M(B^{*}_{c})$.} \\ 
4$^{1}S_{0}$ & 7614.6 & 7608.9 & 7603.1 & 4$^{3}S_{1}$ & 7614.6 & 7616.5 & 
7618.4 & $m_{c}=1.321~GeV$ \\ 
$1^{1}P_{1}$ & 6740.7 & 6732.9 & 6725.0 & $1^{3}P_{2}$ & 6746.9 & 6773.4 & 
6796.6 & $m_{b}=4.731~GeV$ \\ 
2$^{1}P_{1}$ & 7152.2 & 7146.0 & 7139.7 & 2$^{3}P_{2}$ & 7155.6 & 7175.1 & 
7196.3 &  \\ 
$1^{1}D_{2}$ & 7019.4 & 7013.8 & 7008.2 & $1^{3}D_{3}$ & 7012.2 & 7046.2 & 
7078.8 &  \\ 
2$^{1}D_{2}$ & 7371.0 & 7366.1 & 7361.2 & 2$^{3}D_{3}$ & 7366.5 & 7394.9 & 
7411.0 &  \\ 
$1^{3}P_{1}$ & 6734.5 & 6714.3 & 6693.8 & $1^{3}P_{0}$ & 6677.3 & 6608.7 & 
6525.4 &  \\ 
2$^{3}P_{1}$ & 7151.8 & 7136.3 & 7120.5 & 2$^{3}P_{0}$ & 7132.4 & 7091.6 & 
7049.0 &  \\ 
$1^{3}D_{2}$ & 7024.1 & 7010.5 & 6996.8 & $1^{3}D_{1}$ & 7024.6 & 6973.8 & 
6921.0 &  \\ 
2$^{3}D_{2}$ & 7375.2 & 7363.2 & 7351.2 & 2$^{3}D_{1}$ & 7376.1 & 7332.2 & 
7286.9 & 
\end{tabular}
\end{table}

\bigskip \widetext
\begin{table}[tbp]
\caption{$c\overline{c}$ and $b\overline{b}$ mass spectra (in $MeV$) using
the Cornell potential. }
\label{table12}
\begin{tabular}{llllllllllll}
State & Meson\tablenote{Same parameter fits as in Table VI.} & Theory [45] & 
$\epsilon =1$ & $\epsilon =1/2$ & $\epsilon =0$ & Meson & Theory [45] & $%
\epsilon =1$ & $\epsilon =1/2$ & $\epsilon =0$ &  \\ 
\tableline1$^{1}S_{0}$ & $\eta _{c}$ & 2979 & 3068.1 & 3020.8 & 2970.7 & $%
\eta _{b}$ & 9400 & 9448.3 & 9441.3 & 9434.2 &  \\ 
2$^{1}S_{0}$ & $\eta _{c}^{\prime }$ & 3588 & 3686.7 & 3659.6 & 3631.9 & $%
\eta _{b}^{\prime }$ & 9993 & 10021.5 & 10018.3 & 10015.0 &  \\ 
3$^{1}S_{0}$ & $\eta _{c}^{\prime \prime }$ & 3991 & 4135.4 & 4115.0 & 4094.2
& $\eta _{b}^{\prime \prime }$ & 10328 & 10380.4 & 10378.1 & 10375.9 &  \\ 
4$^{1}S_{0}$ &  &  & 4510.7 & 4493.8 & 4476.6 &  &  & 10669.8 & 10668.0 & 
10666.2 &  \\ 
$1^{1}P_{1}$ & $h_{c}$ & 3526 & 3505.7 & 3482.7 & 3459.3 & $h_{b}$ & 9901 & 
9901.3 & 9898.4 & 9895.5 &  \\ 
2$^{1}P_{1}$ & $h_{c}^{\prime }$ & 3945 & 3974.1 & 3955.5 & 3936.7 & $%
h_{b}^{\prime }$ & 10261 & 10266.0 & 10263.9 & 10261.7 &  \\ 
$1^{1}D_{2}$ &  & 3811 & 3812.7 & 3796.0 & 3779.1 &  & 10158 & 10150.5 & 
10148.5 & 10146.5 &  \\ 
2$^{1}D_{2}$ &  &  & 4218.2 & 4203.6 & 4188.8 &  &  & 10456.4 & 10454.8 & 
10453.1 &  \\ 
$1^{3}P_{1}$ & $\chi _{c1}$ & 3510 & 3500.6 & 3463.7 & 3425.7 & $\chi _{b1}$
& 9892 & 9894.9 & 9890.5 & 9886.0 &  \\ 
2$^{3}P_{1}$ & $\chi _{c1}^{\prime }$ & 3929 & 3975.5 & 3946.9 & 3917.8 & $%
\chi _{b1}^{\prime }$ & 10255 & 10263.7 & 10260.5 & 10257.2 &  \\ 
$1^{3}D_{2}$ &  & 3813 & 3820.8 & 3795.6 & 3770.0 &  & 10158 & 10150.1 & 
10147.1 & 10144.1 &  \\ 
2$^{3}D_{2}$ &  &  & 4225.4 & 4203.2 & 4180.9 &  &  & 10456.3 & 10453.8 & 
10451.3 &  \\ 
1$^{3}S_{1}$ & $J/\psi $ & 3096 & 3068.1 & 3083.3 & 3098.2 & $\Upsilon $ & 
9460 & 9448.3 & 9450.6 & 9452.9 &  \\ 
2$^{3}S_{1}$ & $\psi ^{\prime }$ & 3686 & 3686.7 & 3695.6 & 3704.4 & $%
\Upsilon ^{\prime }$ & 10023 & 10021.5 & 10022.6 & 10023.7 &  \\ 
3$^{3}S_{1}$ & $\psi ^{\prime \prime }$ & 4088 & 4135.4 & 4142.1 & 4148.8 & $%
\Upsilon ^{\prime \prime }$ & 10355 & 10380.4 & 10381.2 & 10381.9 &  \\ 
4$^{3}S_{1}$ & $\psi ^{\prime \prime \prime }$ &  & 4510.7 & 4516.3 & 4521.9
& $\Upsilon ^{\prime \prime \prime }$ &  & 10669.8 & 10670.4 & 10671.0 &  \\ 
$1^{3}P_{2}$ & $\chi _{c2}$ & 3556 & 3517.1 & 3566.0 & 3613.5 & $\chi _{b2}$
& 9913 & 9910.3 & 9916.7 & 9923.0 &  \\ 
2$^{3}P_{2}$ & $\chi _{c2}^{\prime }$ & 3972 & 3978.8 & 4019.3 & 4059.0 & $%
\chi _{b2}^{\prime }$ & 10268 & 10269.9 & 10274.7 & 10279.5 &  \\ 
$1^{3}D_{3}$ &  & 3815 & 3802.8 & 3871.4 & 3937.8 &  & 10162 & 10153.1 & 
10161.5 & 10169.8 &  \\ 
2$^{3}D_{3}$ &  &  & 4210.1 & 4270.6 & 4329.5 &  &  & 10458.2 & 10465.1 & 
10472.1 &  \\ 
$1^{3}P_{0}$ & $\chi _{c0}$ & 3424 & 3369.4 & 3106.0 & - & $\chi _{b0}$ & 
9863 & 9864.0 & 9850.7 & 9837.2 &  \\ 
2$^{3}P_{0}$ & $\chi _{c0}^{\prime }$ & 3854 & 3937.0 & 3849.1 & 3752.9 & $%
\chi _{b0}^{\prime }$ & 10234 & 10251.9 & 10243.0 & 10234.0 &  \\ 
$1^{3}D_{1}$ &  & 3798 & 3817.8 & 3711.0 & 3594.8 &  & 10153 & 10144.3 & 
10132.4 & 10120.4 &  \\ 
2$^{3}D_{1}$ &  &  & 4224.0 & 4132.6 & 4035.6 &  &  & 10452.3 & 10442.5 & 
10432.7 & 
\end{tabular}
\end{table}

\bigskip \mediumtext

\bigskip 
\begin{table}[tbp]
\caption{Pseudoscalar and vector decay constants $%
(f_{P}=f_{B_{c}},~f_{V}=f_{B_{c}^{\ast }})$ of the $B_{c}$ meson (in $MeV$). 
}
\label{table13}
\begin{tabular}{llllllll}
Constants & SLNET\tablenote{We used different potential models with
($\epsilon=0$).} & rel [45] & NR [45] & [23] & [29] & [25] & [48] \\ 
\tableline$f_{B_{c}}$ & 378-513 & 433 & 562 & 479-687 & 460$\pm 60$ & 517 & 
420$\pm 13$ \\ 
$f_{B_{c}^{\ast }}$ & 347-482 & 503 & 562 & 479-687 & 460$\pm 60$ & 517 & 
\end{tabular}
\end{table}

\bigskip \mediumtext

\bigskip 
\begin{table}[tbp]
\caption{The radial wave function at the origin (in $GeV^{3})$ calculated in
our model and by the other authors.}
\label{table14}
\begin{tabular}{llllll}
Level & SLNET\tablenote{We used different potential models with
($\epsilon=0$).} & Martin & EQ[23] & F[25] & [23]\tablenote{For the 1S
level.} \\ 
\tableline$\left| R_{B_{c}}(0)\right| ^{2}$ & 0.94-1.73 & 1.716 & 1.638 & 
1.81 & 1.508-3.102 \\ 
$\left| R_{B_{c}^{\ast }}(0)\right| ^{2}$ & 0.73-1.52 &  &  &  & 
\end{tabular}
\end{table}

\bigskip

\bigskip

\bigskip

%\end{document}

\end{document}